\documentclass[conference]{IEEEtran}  % Comment this line out
\let\labelindent\relax         % prevents a conflict with IEEE \labelindent when using enumitem

\usepackage{times}
\usepackage{graphicx}\graphicspath{{fig/}} 
\usepackage{epstopdf}
\usepackage{algorithm}
\usepackage{algorithmic}
\usepackage{multirow} 
\usepackage{multicol}
\usepackage{array}
\usepackage{mdwmath}
\usepackage{mdwtab}
\usepackage{rotating}
\usepackage{caption}
\usepackage{amssymb}
\usepackage{verbatim}
\usepackage{isomath}
\usepackage{overpic}
\usepackage[list=off]{caption}

\makeatletter\newcommand{\manuallabel}[2]{\def\@currentlabel{#2}\label{#1}}\makeatother

\usepackage{color}
\usepackage{xcolor}

\colorlet   {lightorange}{orange!20}
\colorlet   {lightgrey}  {gray!20}

\newcommand{\newabbreviation}[2]{#1\ (\renewcommand{#1}{#2}#1)}

\newcommand{\VIA    }{variable impedance actuator}

\newcommand{\VSA    }{variable stiffness actuator}

\newcommand{\VIAs   }  {\VIA s}
\newcommand{\MACCEPA}  {Mechanically Adjustable Compliance and Controllable Equilibrium Position Actuator}
\newcommand{\MACCEPAVD}{\MACCEPA\ with Variable Damping}
\newcommand{\ILQR   }  {Iterative Linear Quadratic Regulator}
\newcommand{\VSAs}     {\VSA s}
\usepackage{amssymb}
\usepackage{mathtools}

\mathchardef\mhyphen="2D   % define "math hyphen"

\newcommand{\T}     {\top}                % transpose
\newcommand{\intd}  {\mathrm{d}}          % variable of integration
\newcommand{\bx}     {\mathbf{x}}         % (bold) state
\newcommand{\qdot}   {\dot{q}}            % joint velocities
\renewcommand{\nu}  {q}                   % index for elements of u
\newcommand  {\dt}         {\,\intd t}             % t as variable of integration

\newcommand{\inertia}{m} % inertia

\newcommand{\DCr}{D_r}
\newcommand{\DCd}{D_d}       % symbol definitions
\graphicspath{
{figures/}
}
\providecommand{\figurename}{Fig.}
\newcommand*{\sref}[1]{\S\ref{s:#1}}            % section
\newcommand*{\fref}[1]{\figurename~\ref{f:#1}}  % figure
\newcommand*{\eref}[1]{(\ref{e:#1})}            % equation

\usepackage[inline]{enumitem}
\setlist{nolistsep}
\newcommand{\il}[1]{\begin{enumerate*}[label=(\roman*)]#1\end{enumerate*}}

\newcommand{\eg}{\textit{e.g.,}~} %
\newcommand{\ie}{\textit{i.e.,}~} %
\newcommand{\cf}{\textit{cf.}~} %
\newcommand*{\schemeref}[1]{Scheme \ref{scheme:#1}} % referencing format
\usepackage[normalem]{ulem}                                        % <- for striking text out
\usepackage{marginnote}
\newcommand{\tinytodo}[2][]
{\todo[caption={#2}, size=\small, #1]{\renewcommand{\baselinestretch}{0.5}\selectfont#2\par}}

\colorlet{jb}{red}
\colorlet{mh}{red}
\colorlet{fw}{green}

\newcommand  {\done}[1]{\sout{#1}}

\newcommand  {\mh}  [1]{\tinytodo[color=white,linecolor=mh,bordercolor=white,noinline]{\protect{\scriptsize\color{mh}#1}}}
\newcommand  {\fw}  [1]{\tinytodo[color=white,linecolor=fw,bordercolor=white,noinline]{\protect{\scriptsize\color{fw}#1}}}
\usepackage[disable]{todonotes}%\newcommand{\mh}[3]{#3}{}{}\newcommand{\maybedo}[1]{}

\newcommand{\atFW} {{\color{fw}@FW}}

\usepackage{amsfonts}
\usepackage{amsmath}
\usepackage{amssymb}
\usepackage{graphicx}
\usepackage{caption}
\usepackage{subcaption}

\IEEEoverridecommandlockouts
\title{\LARGE \bf%
A Hybrid Dynamic-regenerative Damping Scheme for Energy Regeneration in Variable Impedance Actuators
}

\author{Fan Wu and Matthew Howard$^{*\dagger}$ % <-this % stops a space
\thanks{$^{*}$Fan Wu and Matthew Howard are with the Centre for Robotics Research, Department of Informatics, King's College London, UK {\tt\small \{fan.wu, matthew.j.howard\}@kcl.ac.uk.}}%
\thanks{$^{\dagger}$This work was supported in part by the UK Engineering and Physical Sciences Research Council (EPSRC) SoftSkills project, EP/P010202/1.}%
}
\renewcommand{\baselinestretch}{0.905}
\begin{document}
%\usage
%\input{tex/usage}%\clearpage\listoftodos
\maketitle

\begin{abstract}
	Increasing research efforts have been made to improve the energy efficiency of variable impedance actuators (VIAs) through reduction of energy consumption. However, the \emph{harvesting} of dissipated energy in such systems remains under-explored. This study proposes a novel variable damping module design enabling energy regeneration in VIAs by exploiting the regenerative braking effect of DC motors. The proposed damping module uses four switches to combine \emph{regenerative} and \emph{dynamic} braking, in a hybrid approach that enables energy regeneration without reduction in the range of damping achievable. Numerical simulations and a physical experiment are presented in which the proposed module shows an optimal trade-off between task-performance and energy efficiency.%It ensures the current generated from bidirectional rotation remains unidirectional for charging electric storage element. 
	%The application on VIAs is evaluated to show the advantage over dynamic and  pure regenerative damping, which improves the energy efficiency and also maintains maximum achievable damping for optimal performance-energy trade-off balancing.
\end{abstract}
%Optimal control has been increasingly applied to exploit control redundancy of \VIAs. Energy efficiency is a crucial issue for both mechanical design and control. From the optimal control perspective, optimisation with minimal principle concerning energy cost of \VIA s has been limited to smoothness criterion. A more plausible cost estimation is proposed to optimise \VIA s trajectories for energy-optimal task execution. Furthermore, the impact of introducing energy regeneration on variable damping component on energy-optimal behaviour is addressed.
%\mh{\done{\atFW: Update the abstract once the rest of the paper is complete.}}
%\mh{\done{\atFW: Please revise the abstract according to your latest findings.}}
%\mh{\done{Avoid using active voice.}} 
%%%%%%%%%%%%%%%%%%%%%%%%%%%%%%%%%%%%%%%%%%%%%%%%%%%%%%%%%%%%%%%%%%%%%%%%%%%%%%%%
\section{Introduction}\label{s:introduction}\noindent
Variable impedance actuators (\renewcommand{\VIA}{VIA}\VIAs) are believed to be the key for the next generation of robots to interact safely with uncertain environments and provide better performance in cyclic tasks and dynamical movements \cite{Vanderborght2012}. For example, the physical compliance incorporated in \emph{\VSAs} (\renewcommand{\VSA}{VSA}\VSAs) (\eg using elastic components such as springs) enables energy storage, which can \il{\item absorb external energy introduced into the system (\eg from collisions) to enhance safety, and \item amplify output power by exploiting the stored energy \cite{Grebenstein2011}}. 

Recently, much research effort has gone into the design of \emph{variable physical damping} actuation, based on different principles of damping force generation (see \cite{Vanderborght2013} for a review). %, to compensate the under-damped oscillations and facilitate better exploitation of their natural dynamics for performance enhancement in highly dynamic tasks. 
Variable physical damping has proven to be necessary to achieve better task performance when higher damping is desired to compensate undesired oscillations caused by the use of physical stiffness \cite{Laffranchi2012a, Laffranchi2013a}. 
It is also demonstrated in \cite{Laffranchi2012a} that variable physical damping plays an important role in terms of energy efficiency for actuators with fixed stiffness or a small range of variable stiffness, to optimally adjust dynamic properties. 
However, while these studies represent important advances in terms of improving the efficiency of \emph{energy consumption} in \VIAs, the importance of variable physical damping may be underestimated, because the potential to \emph{harvest energy dissipated by damping} has so far received little attention.

\begin{figure}[t!]
	\centering%
	\includegraphics[width=\linewidth]{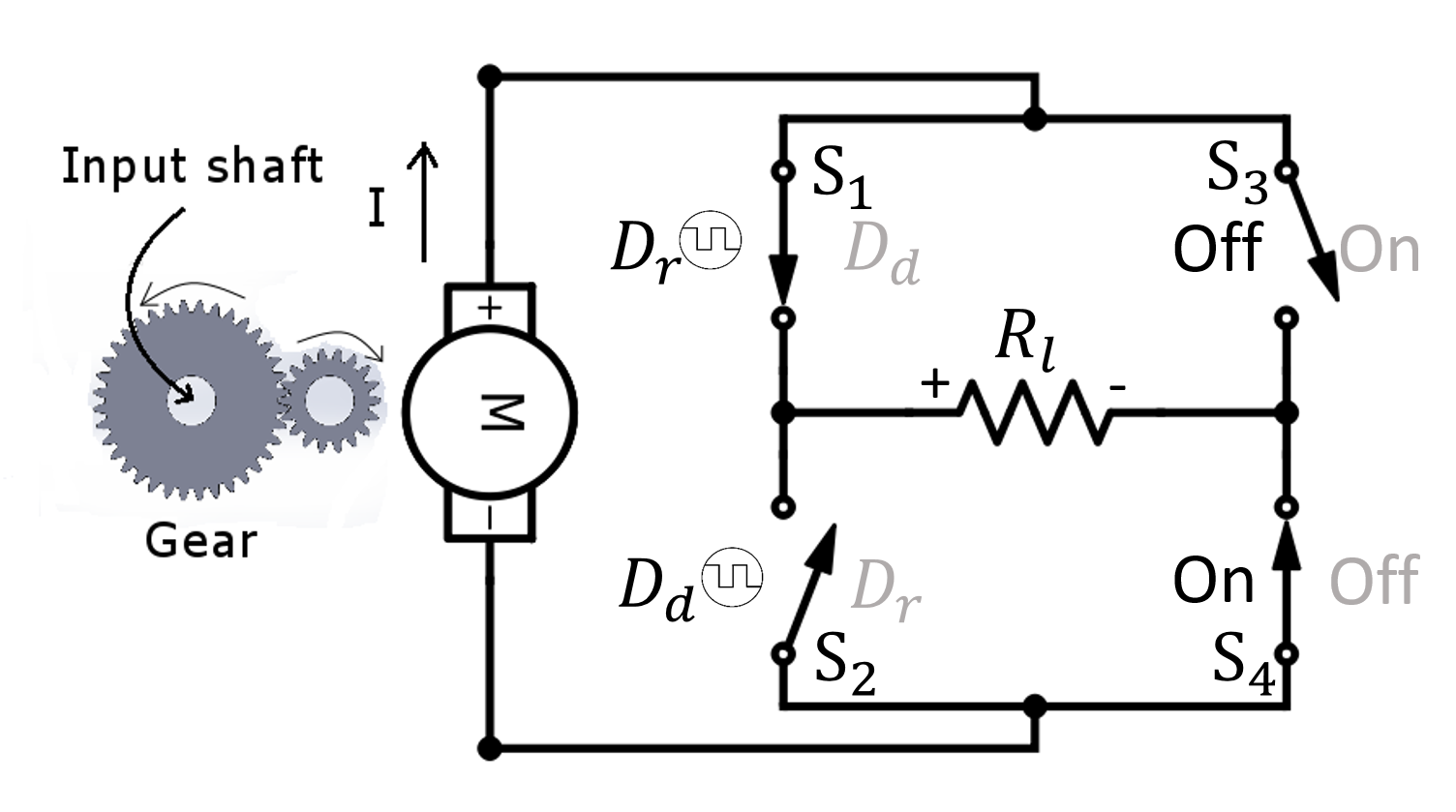}
	\caption{Circuit diagram for bidirectional hybrid dynamic-regenerative braking.}
	\label{f:scheme-bidirection}
\end{figure}
\mh{\done{\atFW: Update \fref{switch-bidirection} to include the different schemes.}}
To address this, this paper extends the variable damping technique introduced by \cite{Radulescu2012} to take into account of the energy regeneration capabilities of DC motors. 
In contrast to prior approaches that incorporate a power source directly (as done, for example, in \cite{Donelan2008}), a circuit design using four switches---considering the fact that the rotation of a revolute robot joint is bidirectional while typical power sources for energy storage are unidirectional---is introduced that enables adjustment of the electrical damping effect, while increasing the damping range available to the controller.
The relation of the damping effect and the power of regeneration of the proposed damping module is investigated, and shows a non-monotonic relation emerges that requires balancing the trade-off between damping allocation and energy regeneration in a non-trivial way. Nevertheless, the proposed controllable damping module is evaluated in terms of movement performance and energy recovery of a simple reaching task on a ideal pendulum model and a more realistic \VIA\ model, and shows that an optimal trade-off is achievable. 
\mh{\done{\atFW: ``an optimal trade-off is achievable.''?}} 
Furthermore, an experiment is presented in which the damping module is realised in hardware, verifying the theoretical predictions about the damping module's behaviour.
%
%It is further realised in hardware, whereby the electronics design is implemented based on the proposed method. 
%The experiment for characterising the damping effect and the power of regeneration verifies the non-monotonic relation between them and demonstrates the efficacy of proposed variable hybrid damping module design.
%It is further verified in an evaluation in hardware, whereby the hybrid damping module and its electronics design is realised according to the proposed method. 
\mh{\done{\atFW: say something quantitative about the hardware results here, showing the improvement over pure dynamic/regenerative braking.}}

\section{Background}\label{s:background}\noindent
%\subsection{Variable damping with energy regeneration}\noindent \label{s:models}
%\section{Design of variable damping with energy regeneration}\noindent \label{s:models}
To date, two main approaches have been proposed that enable variable damping through the use of DC motors, namely \il{\item \emph{dynamic braking} and \item \emph{regenerative braking}}. In both cases, the back electromotive force is used to resist movement proportional to the effective resistance of the damper motor circuit, causing a variable damping effect. The following briefly outlines these schemes.
\begin{figure*}[t]
\begin{overpic}[width=0.32\linewidth]{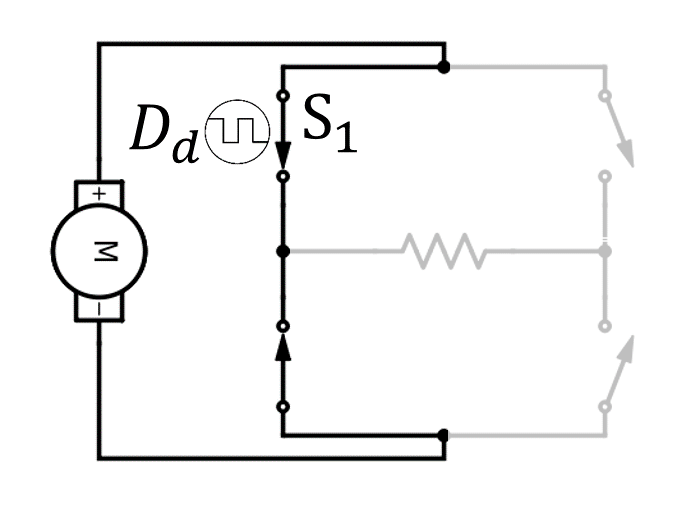}\put(0,5){\ref{f:scheme-dynamic}}\end{overpic}%
\begin{overpic}[width=0.32\linewidth]{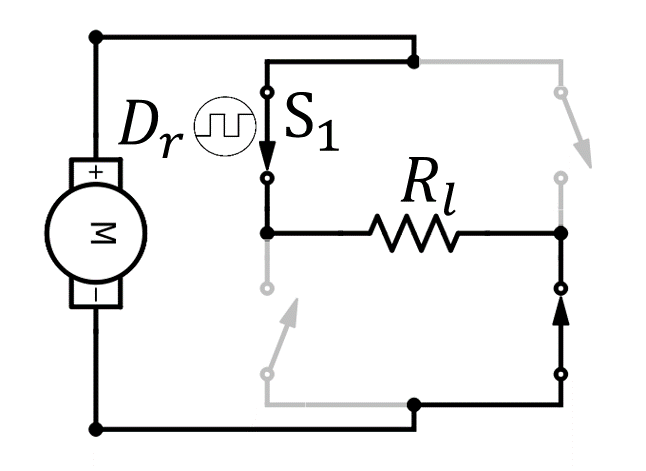}\put(0,5){\ref{f:scheme-rege}}\end{overpic}%
\begin{overpic}[width=0.32\linewidth]{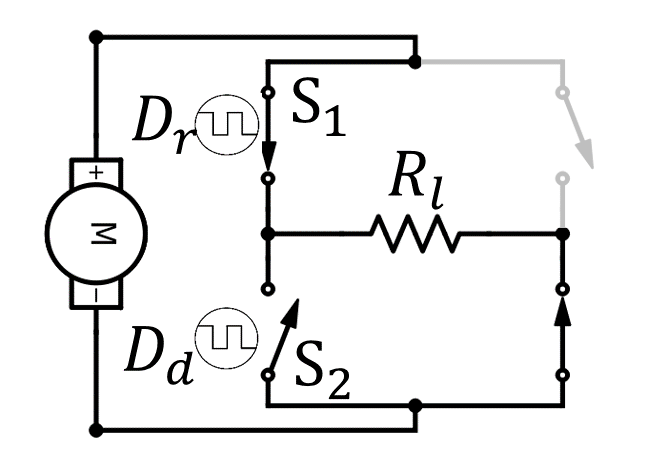}\put(0,5){\ref{f:scheme-hybrid}}\end{overpic}%
	\caption{\label{f:schemes}Conceptual diagrams of 
\begin{enumerate*}[label=(\alph*)]
\item\label{f:scheme-dynamic} dynamic, 
\item\label{f:scheme-rege} regenerative, and
\item\label{f:scheme-hybrid} hybrid dynamic-regenerative
\end{enumerate*} 
 braking circuits.
    }
\end{figure*}
\mh{\done{\atFW: Please make the grey parts of \fref{schemes}\ref{f:scheme-dynamic}-\ref{f:scheme-hybrid} even lighter.}}
\mh{\done{\atFW: Please make the top right part of \fref{schemes}\ref{f:scheme-hybrid} grey, since it does not play a role in unidirectional hybrid braking.}}
%\footnotetext{Another somehow dangerous case is \emph{concurrent braking} when R also works as source and contributes to build up $\tau_m$, resulting in large current and huge heat dissipation on the circuit.}
\subsubsection*{\schemeref{dynamic-braking} - Dynamic braking}\manuallabel{scheme:dynamic-braking}{1}
Dynamic braking in the context of \VIA\ design was first proposed by \cite{Radulescu2012}. A circuit diagram for this scheme is illustrated in \fref{schemes}\ref{f:scheme-dynamic}. 
%\mh{\atFW: Please create \fref{scheme-dynamic}.} 
In this mode, the damping effect is modulated by changing the duty-cycle $\DCd$ that controls the portion of time that a switch $S_1$ spends in the open or closed position, thereby altering the effective resistance of the circuit. The damping coefficient follows the equation
\begin{equation}
	d = \frac{n_d^2 k_t^2 \DCd}{R_m}= \bar{d}_1 D_d \label{e:d-scheme-dynamic}
\end{equation} 
where $n_d$ is the gear ratio of damping motor, $k_t$ is the torque constant and $R_m$ is the resistance of the motor. Note that, since $0\le \DCd \le1$, the maximum damping coefficient that can be provided by dynamic braking is $\bar{d}_1=n_d^2 k_t^2/R_m$. 

In energy terms, dynamic braking is effective %\fw{ineffective?}\mh{\atFW: effective---in the sense that it works to damp the movement} 
since it dissipates kinetic energy of the output link as heat in the electrical circuit. It does not, however, charge energy to any electrical source, so the regeneration power is zero ($P_\mathrm{rege} = 0$). In other words, this (potentially useful) energy is simply discarded, reducing the overall energy efficiency of the system.

\subsubsection*{\schemeref{regenerative-braking} - Regenerative braking}\manuallabel{scheme:regenerative-braking}{2} 
Regenerative braking refers to the situation where the power generated by the motor through kinetic motion of the output link is used to recharge an electrical storage element \mh{\done{\atFW: Is ``electrical source'' the right word for this? Something like ``electrical storage element'' would seem a better term.}} (\eg battery, supercapacitor). To implement regenerative braking, the electrical storage element can be simply connected to the circuit of the damping motor, as shown in \cite{Donelan2008}. In the context of \VIA\ design, this can be implemented through the circuit in \fref{schemes}\ref{f:scheme-rege}. %\mh{\atFW: Please create \fref{Fig1b}.} 

In regenerative braking mode, the damping effect is dependent on the combined effective resistance of the circuit containing the electrical storage element. Similar to dynamic braking, this can be modulated by controlling the duty-cycle $\DCr$ of a switch. The damping coefficient and the regeneration power can be calculated as
\begin{align}
	d &= \frac{n_d^2 k_t^2 \DCr}{R_m + R_l} = \bar{d}_2 \DCr \label{e:d-scheme-rege}\\
	P_\mathrm{rege} &= \frac{R_l n^2_d k^2_b \dot{q}^2 D_r}{(R_m + R_l)^2}=\alpha \bar{d}_2 \dot{q}^2 D_r, \label{e:Prege-scheme-rege}
\end{align}
respectively, where $R_l$ is the resistance of the electrical source and $\alpha = R_l/(R_m+R_l)$, $k_b$ is back-EMF constant and equals to $k_t$.

Note that, introducing regenerative braking means that the mechanical energy that is otherwise discarded in the dynamic braking scheme can be harvested, enhancing the overall energy efficiency of the system. However, note also that, compared to dynamic braking, the maximum damping coefficient that can be produced by regenerative braking, $\bar{d}_2=n_d^2 k_t^2/(R_m+R_l)$, is \emph{decreased} since adding electrical load for charging increases the total equivalent resistance of the circuit. This can be a drawback in applications where higher levels of damping are needed (\eg when there is need for a high dynamic response and therefore heavy braking of rapid movements). 

\section{Hybrid dynamic-regenerative braking}\noindent \label{s:hybrid-damping}
To address these issues, this paper proposes a variable damping scheme---termed \emph{hybrid braking}---that switches between dynamic braking and pure regenerative braking to achieve the optimal benefits of both. 

\subsection{Hybrid damping circuit}\noindent
The hybrid damping scheme is implemented through the circuit depicted in \fref{schemes}\ref{f:scheme-hybrid}.
\label{s:energy-rege}
It uses two switches (denoted $S_i$, $i\!\in\!\{1,2\}$) that switch at high frequency between \il{\item pure regenerative braking, and \item a blend of dynamic and regenerative braking}. The principle by which the proposed scheme operates is as follows. 

When switch $S_2$ is open, the module acts in regenerative braking mode, whereby current flows through the power storage element, with the effective resistance (damping level) determined by the duty cycle of $S_1$. (Note that, this results in an equivalent circuit to that used in \schemeref{regenerative-braking}, \cf \fref{schemes}\ref{f:scheme-rege}.) On the other hand, when $S_1$ and $S_2$ are closed, there is a short circuit that causes current to bypass the resistive load $R_l$, creating a dynamic braking effect. In this case, the damping level can be determined by keeping $S_1$ closed and modulating the duty cycle of $S_2$. This enables a third braking scheme to be defined, alongside Schemes \ref{scheme:dynamic-braking} and \ref{scheme:regenerative-braking},
as follows.

\subsubsection*{\schemeref{hybrid-braking} - Hybrid braking}\manuallabel{scheme:hybrid-braking}{3} 
When the required damping $d^*$ is small enough, \ie $d^*\le \bar{d}_2$, 
\mh{\done{\atFW: $d_{\mathrm{rege}}^{max}=\bar{d}_1$?}} 
it can be provided by pure regenerative braking, so $S_2$ is opened ($\DCd=0$).  
When the required damping is greater, \ie $d^*>\bar{d}_2$, $S_1$ is closed ($\DCr=1$) and $\DCd$ is used to control $S_2$ to blend dynamic and regenerative braking. 

The resulting damping coefficient and regeneration power are:
\begin{align}
	d &= \bar{d}_2 \DCr + \alpha \bar{d}_3 \DCd \label{e:d-scheme-hybrid0}\\
	P_\mathrm{rege} &= \alpha \bar{d}_2 \dot{q}^2 (\DCr - \DCd).\label{e:Prege-scheme-hybrid}
\end{align}
Note that, if $\DCr=\DCd=1$, the same maximum damping coefficient as that achievable in a pure dynamic braking can be achieved, \ie $\bar{d}_3=\bar{d}_1$. This, however, comes at the cost of the regeneration power vanishing ($P_\mathrm{rege}=0$).
%\mh{\atMH: Fresh comments to here (7/9/2017).}

\subsection{Hybrid Damping Control Modes}\label{s:control-modes}\noindent
In principle, each of the switches in the proposed circuit may be independently controlled by its own duty-cycle. While this enhances the flexibility of the damping module, it also introduces an undesirable layer of complexity to its control. 

\begin{figure}[t!]
\begin{overpic}[width=0.49\linewidth]{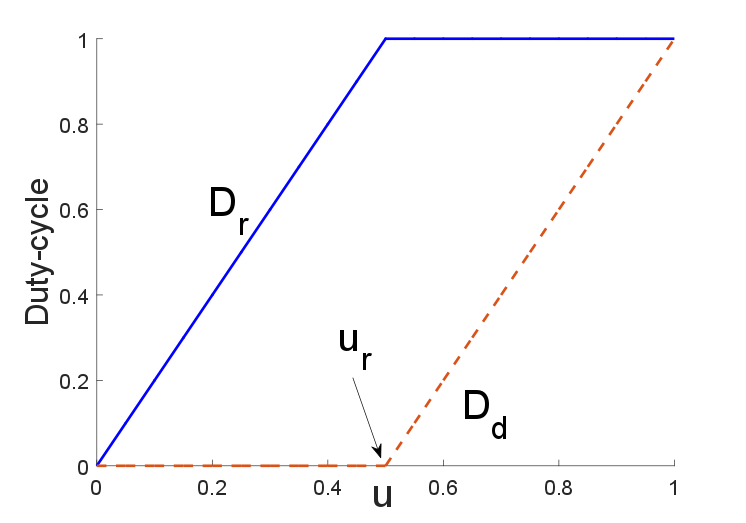}\put(0,5){\ref{f:D1D2-u}}\end{overpic}%
\hfill%
\begin{overpic}[width=0.49\linewidth]{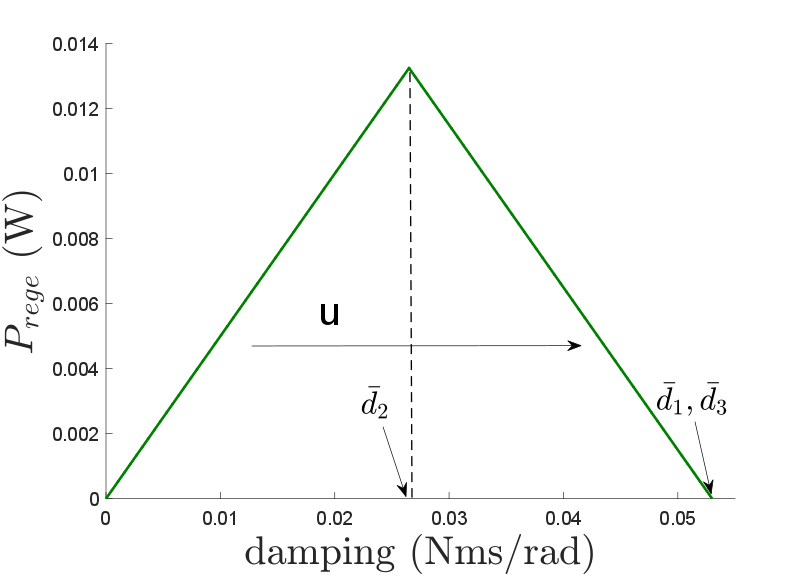}\put(0,5){\ref{f:Prege-damping}}\end{overpic}%
	%\begin{subfigure}{.49\linewidth}
	%\centering
	%\includegraphics[width=\linewidth]{figures/D1D2-u.eps}
	%\caption{Mapping from control input $u$ to duty cycles $D_1, D_2$.}
	%\label{f:D1D2-u}	
	%\end{subfigure}
	%\hfill
	%\begin{subfigure}{.49\linewidth}
	%	\centering
	%	\includegraphics[width=\linewidth]{figures/Prege-damping.eps}
	%	\caption{Relation between power of regeneration and damping.}
	%	\label{f:Prege-damping}
	%\end{subfigure}
	\caption{\label{f:D1D2-u_Prege-damping} Hybrid damping control modes.
\begin{enumerate*}[label=(\alph*)]
\item\label{f:D1D2-u} Mapping from control input $u$ to duty cycles $D_r, D_d$.
\item\label{f:Prege-damping} Relation between regeneration power and damping.
\end{enumerate*}
    }
\end{figure}

To address this, and enable the simple control of the module through a single control variable $u \in [0,1]$, the duty cycles of the switches can be coupled through the following relation
\begin{align}
	D_r &= \left\{
	\begin{aligned}
	\frac{u}{u_r}, u \leqslant u_r\\
	1, u > u_r
	\end{aligned}
	\right. \nonumber \\
	D_d &= \left\{
	\begin{aligned}
	0, u \leqslant u_r\\
	\frac{u - u_r}{1 - u_r}, u > u_r
	\end{aligned}
	\right.\label{e:D1D2}
\end{align}
where $u_r$ corresponds to the maximum of damping coefficient of regenerative braking ($d(u_r) = \bar{d}_2$) and depends on user's selection. 
\mh{\done{\atFW: The definition of $u_r$ is still not clear -- how is the function $d(x)$ defined?}}
In this paper, $u_r$ is chosen to be $0.5$. Substituting \eref{D1D2} into \eref{d-scheme-hybrid0}, damping coefficient $d(u)$ as a function of $u$ is simplified to:
\begin{equation}
	d(u) = \bar{d}_3 u \label{e:d-scheme-hybrid}
\end{equation}
As illustrated in \fref{D1D2-u_Prege-damping}\ref{f:D1D2-u}, 
\mh{\done{\atFW: Mark $u_r$ in \fref{D1D2-u_Prege-damping}\ref{f:D1D2-u}. Strictly limit the range of the vertical axis between zero and one. Label the vertical axis ``Duty Cycle''.}}
when $u\leqslant 0.5$, $D_d$ remains at zero (\ie switch $S_2$ is open) and $D_r$ is linearly mapped from $u \in [0,0.5]$ to $[0,1]$, while when $u > 0.5$, $D_r$ is held at unity ($D_r=1$ so $S_1$ is closed) and $D_d$ is linearly mapped from $u \in [0.5,1]$ to $[0,1]$\mh{\done{\atFW: Define $u_r$.}}. 
\mh{\done{\atFW: Explain \eref{D1D2}, and justify its use. Refer to \fref{D1D2-u}.} }

The relation between the damping coefficient $d$ and the power regeneration $P_\mathrm{rege}$ for a fixed angular velocity is shown in \fref{D1D2-u_Prege-damping}\ref{f:Prege-damping}. As can be seen, the relationship is non-monotonic and there is a peak for $P_\mathrm{rege}$ when $d = \bar{d}_2$, \ie at the upper boundary of the pure regenerative braking domain.
\mh{\done{\atFW: Axis labels in \fref{Prege-damping}.}}
\mh{\atFW: Out of curiosity -- maybe it would be beneficial to design a smooth function for this -- especially if it is to be used in combination with an optimiser like ILQR.}
    \mh{\done{\atFW: Create separate diagrams for Schemes 1-3 in \fref{D1D2-u}, plus one for the bidirectional version of Scheme 3. The diagrams can placed side-by-side in one figure.}}
    %\mh{\atFW: Consider overlaying \fref{Prege-damping} on the \emph{measured} power-damping curve.}
%\mh{\atFW: Create \fref{scheme-hybrid}.} 

\subsection{Bidirectional damping}\noindent
The hybrid damping circuit described so far enables the modulation of damping force associated with unidirectional motion of the output link. In order to realise damping of bidirectional motion (as is common in many robotic applications), it is necessary to ensure that the current generated by the damping motor always flows into the positive terminal of the electrical source. This can be achieved by a four-switch design of the damping circuit, as illustrated in \fref{scheme-bidirection}.
When the current flows from the positive terminal of the damping motor (as shown by the black arrow in \fref{scheme-bidirection}), $S_3$ is open and $S_4$ is switched on. When the current flows from the negative terminal of the motor (as shown by the grey arrow), $S_3$ is closed and $S_4$ is open, and $S_1$ is controlled by $D_d$ and $S_2$ is controlled by $D_r$. 
\mh{\done{\atFW: Briefly explain how this solves the problem.}}
\mh{\done{\atFW: This last sentence is quite confusing. It may be better to use different symbols \ie instead of $D_1$, $D_2$, use $D_r$, $D_d$ -- since these control the regenerative and dynamic parts of the hybrid scheme. Note that, you will need to update this in your figures as well as the text.}}

It should be further noted that, this latter circuit, implements the (bidirectional versions of) the two damping schemes outlined in \sref{background} as special cases. For example, \il{\item holding $S_2,S_3$ open, $S_4$ closed and varying the duty cycle of $S_1$ results in regenerative braking, while \item holding $S_3,S_4$ open, $S_1$ closed, and varying the duty cycle of $S_2$ results in pure dynamic braking}. In other words, the same hardware can be used to realise all three damping schemes.

%\subsubsection*{Scheme 1 - Dynamic braking}
%In this mode, the electrical source is always disconnected. Hence, $S_2$ keeps on all the time ($D_2 = 1$). 
%
%The damping coefficient is modulated by changing the duty-cycle $D_1$, which follows the equation:
%\begin{equation}
%	d = \frac{n_d^2 k_t^2 D_1}{R_m}= \bar{d}_0 D_1 \label{e:d-scheme0}
%\end{equation} 
%where $n_d$ is the gear ratio of damping motor, $k_t$ is the torque constant, $R_m$ is the resistance of the motor, $\bar{d}_0=n_d^2 k_t^2/R_m$ is the maximum damping coefficient it can provide by dynamic braking. Since it does not charge energy to electrical source in this mode, the power of regeneration is $0$ ($P_\mathrm{rege} = 0$).
%
%\subsubsection{Scheme 2 - Regenerative braking}
%In this pure regenerative braking mode, the switch $S_2$ is off all the time ($D_2 = 0$) and $D_1$ is the control input. The damping coefficient and the power of regeneration can be calculated by:
%\begin{align}
%	d &= \frac{n_d^2 k_t^2 D_1}{R_m + R_l} = \bar{d}_1 D_1 \label{e:d-scheme1}\\
%	P_\mathrm{rege} &= \frac{R_l n^2_d k^2_b \dot{q}^2 D_1}{(R_m + R_l)^2}=\alpha \bar{d}_1 \dot{q}^2 D_1 \label{e:Prege-scheme1}
%\end{align}
%where $\bar{d}_1=n_d^2 k_t^2/(R_m+R_l)$ is the maximum damping coefficient it can produce by regenerative braking. $\alpha = R_l/(R_m+R_l)$ is determined by the ratio between $R_l$ and $R_m$.

\section{Evaluation}\label{s:evaluation}\noindent
This section evaluates the proposed hybrid braking scheme in comparison to pure dynamic or regenerative braking through numerical simulation
of \il{\item a simple pendulum actuated with an ideal \VIA, and \item a more realistic simulation of a physical \VIA, namely the \newabbreviation{\MACCEPAVD}{MACCEPA-VD} \cite{Radulescu2012,VanHam2007}}.

%introduced above
%to show that \il{\item how much energy it can be regenerated via regenerative and hybrid braking; \item hybrid braking scheme is necessary to eliminate the side effects of regenerative braking.}

\subsection{Simple pendulum with ideal VIA}\label{s:toy-example}\noindent
The aim of the first evaluation is to illustrate the effectiveness of the hybrid braking scheme in the context of a simple example task of target reaching.

For this, a model of a simple pendulum, subject to viscous friction and actuated by an ideal \VIA \footnote{It is assumed that in the VIA model the damper is in parallel with the spring.}, is used
\begin{align}
	m l^2 \ddot{q} + b \dot{q} = k(u_2)( u_1 - q) - d(u_3).
    \label{e:simple-pendulum}
\end{align}
Here, for simplicity, $m=1 \mathrm{kg}, l = 1 \mathrm{m}$, $b=0.01 \mathrm{Nms/rad}$. The right hand side of \eref{simple-pendulum} is the joint torque applied by the ideal \VIA, $u_1 \in [-\pi/2,\pi/2]\,\mathrm{rad}$ controls the equilibrium position and the stiffness $k(u_2)$ is proportional to the control input $u_2\in[0,1]$:
%\mh{$u_2\in[0,1]$?}
\begin{align}
	k(u_2) = \bar{k}u_2
\end{align}
where $\bar{k}=200\,\mathrm{Nm/rad}$ is the maximum stiffness. The parameters\footnote{These parameters are arbitrarily chosen to give response within a second. Experimentation shows the result is not sensitive to these values.}
 that characterise the variable damping module are selected to be $\bar{d}_3 = 50\,\mathrm{Nms/rad}$, $\bar{d}_2 = 25 \mathrm{Nms/rad}$ and $\alpha = 0.5$.
The control frequency is set to $50\mathrm{Hz}$.

The task is to reach a target $q^* = \pi/3\,\mathrm{rad}$ from the initial position $q = 0\,\mathrm{rad}$ within a finite time $t_f$ as quickly and accurately as possible, while minimising the energy consumption and control effort. This can be described through minimisation of the cost function %\mh{\done{\atFW: Define the cost function used.}}
\begin{multline}
	J = \int_{0}^{t_f} [w_1 (q(t) - q^*)^2 + w_2 ( u_1(t)-q^* )^2 \\
	+ w_3 u_2^2(t) - w_4 P_{\mathrm{rege}}] \dt \label{e:cost-function}
\end{multline}
where $w_1=1000$, $w_2=w_3=1$, $w_4=0.01$ are weighting parameters. These parameters are selected to take account of the different scales of the terms and allow reaching within a second.

To simplify the analysis, in the below, the command for equilibrium position is fixed at $u_1 = \pi/3$, while the commands for stiffness and damping are allowed to vary. The optimal open-loop control sequence  for the latter is computed through the \newabbreviation{\ILQR}{ILQR} method \cite{Li2004} with the proposed hybrid braking scheme, and the resultant trajectory of the system is computed by simulating the execution of the open-loop command using the 4th Order Runge-Kutta method. 
\mh{\done{\atFW: Is this the 4th order Runge-Kutta method?}}

To evaluate the energy efficiency of the proposed approach, the total mechanical work and the total regenerated energy are computed from the resultant trajectories, \ie
\begin{align}
E &= \int_0^{t_f} k(u_1 - q)\cdot \dot{q} \dt\\
E_\mathrm{rege} &= \int_{0}^{t_f} P_\mathrm{rege}(t)\dt,
\label{e:toy-E}
\end{align}
respectively. The net energy cost can be defined as
\begin{equation}
E_\mathrm{net}= E - E_\mathrm{rege}.
\end{equation}
The percentage ratio of energy regeneration
%\footnote{Note that, the ratio defined here is different from the notion of system transmission efficiency which is affected by losses due to frictions.}
\footnote{Note that, for simplicity, it is assumed here that there is $100\%$ kinetic to electric energy transmission efficiency of the DC motor. In practice, losses are likely to occur due to friction and losses in the conversion from the mechanical to the electrical domain.} 
can be computed by
\begin{equation}
	\eta = \frac{E_\mathrm{rege}}{E}.
\end{equation}
\mh{\done{\atFW: What exactly do you show in the bottom right panel of \fref{toy_example}, and how is it computed? Is it \eref{toy-E}? If so, please describe it here, along with the computation of the percentage ratio of energy regeneration.}}

For comparison, the experiment is repeated
with \il{\item pure dynamic braking (\schemeref{dynamic-braking}), \item pure regenerative braking (\schemeref{regenerative-braking}), \item the case where the damping is fixed at the maximum power of regeneration ($d=\bar{d}_2$), and \item\label{i:toy-case-critical} a critically damped system}. In the latter, the stiffness is chosen to be $k=100 \mathrm{Nm/rad}$ and the damping is fixed to $d=20 \mathrm{Nms/rad}$ such that the damping ratio $\zeta=d/2\sqrt{km}=1$. 
\mh{\done{\atFW: How did you choose $k$ and $d$ in case \ref{i:toy-case-critical}? Are these the optimal fixed stiffness and damping levels for this problem? Do they comply with the admissible values for the system?}}
\fw{\done{I tried different choices of k and d and this is not the optimal one. There is a selection where it is very close to the fixed damping one.}}
\mh{\atFW: Ok, let's keep it this way for now. When you are back in London, I would like to see your results for the optimal critical damping case.}

%Comparisons are made by investigating the reaching trajectories with damping coefficients and power of regeneration controlled by different schemes presented in \sref{models} \il{\item dynamic; \item regenerative; \item hybrid mode; and two more cases \item critical damping strategy, \item fixed damping at maximum power of regeneration ($d=\bar{d}_1$)}.
%The optimal open-loop solutions of (i - iii) and (v) are solved by ILQR method \cite{Li2004}. For critical damped reaching the stiffness is chosen to be $k=100 \mathrm{Nm/rad}$ and the damping is fixed to $d=20 \mathrm{Nms/rad}$. 
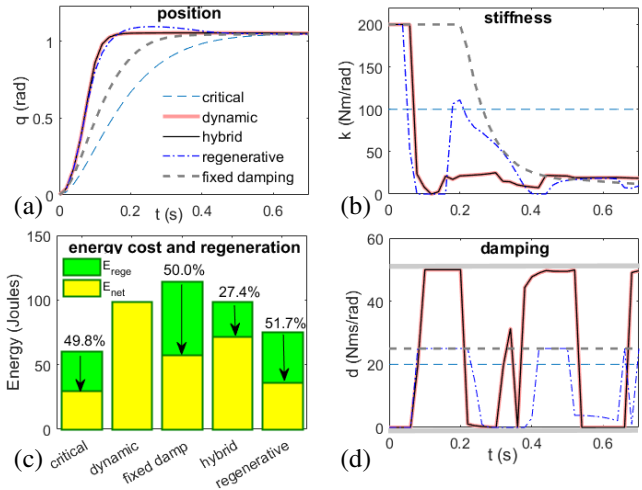
\begin{figure}[t!]
	\centering%
\begin{overpic}[width=\linewidth]{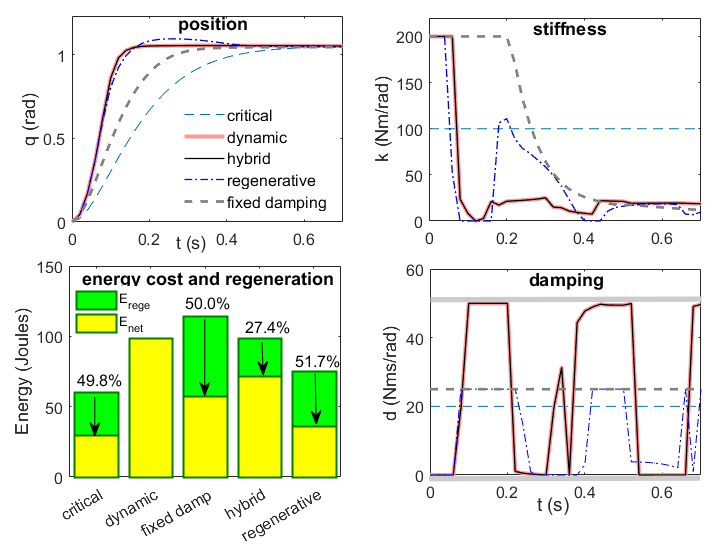}
\put( 3,43){\ref{f:toy-q}}
\put(52,43){\ref{f:toy-k}}
\put( 3, 5){\ref{f:toy-E}}
\put(52, 5){\ref{f:toy-d}}
\end{overpic}%    
%	\includegraphics[width=\linewidth]{figures/toy_example.png}
	%\caption{Test of reaching task on simple pendulum with ideal VIA.}
	%\label{f:toy_example}
	\caption{\label{f:toy_example} Test of reaching task on simple pendulum with ideal VIA. Shown are optimal
\begin{enumerate*}[label=(\alph*)]
\item\label{f:toy-q} joint angular trajectories,
\item\label{f:toy-k} stiffness, and
\item\label{f:toy-E} total mechanical work and percentage ratio of energy regeneration for different damping schemes, and
\item\label{f:toy-d} damping profiles.
\end{enumerate*}
    }
\end{figure}
\mh{\atFW: Figure could be improved by 1) Adding the target angle to the top right panel, 2) showing the admissible range of control values, 3) flipping the green and yellow in the bottom right panel, 4) showing the dynamic and hybrid trajectories laid one on top of the other (use a thicker, lighter shade line for the lower), 5) increasing the fond size for the axis labels, 6) labelling the $y$-axis in the bottom right panel, 7) putting the position and energy plots on the left, and the stiffness and damping plots on the right. 8) Changing 'C.D.' to 'critical'.}
%\fw{modify figure. Dynamic should be a separate traj.}
The results are illustrated in \fref{toy_example}.
%where the position, stiffness and damping of the trajectories are shown in the top two and bottom left figures, respectively, and the bottom right figure shows the energy cost and percentage ratio of regeneration (indicated by the number on each bar).
%
\mh{Shift \eref{toy-E} up to the point indicated above, and justify your use of it.}
As can be seen, the trajectory of the critically damped system reaches the target slowly but without overshoot (\fref{toy_example}\ref{f:toy-q}). The system with fixed damping reaches the target quicker than the critically damped one, because it can exploit the variable stiffness. The system with regenerative braking reaches the target quicker still, however, since the damping range is limited in this case, it suffers from overshoot once it reaches the target. In contrast, the dynamic braking and hybrid braking systems reach the target quickest without overshot, so perform best in terms of accuracy. 

Looking at \fref{toy_example}\ref{f:toy-E}, however, it can be observed that the dynamic braking performs worst in terms of net energy cost, since no energy is recovered throughout the movement. This contrasts with the hybrid approach, that achieves fast and accurate movement while also achieving $27.4\%$ energy recovery, thereby lowering the net energy cost.

%Interestingly, fixing the damping at the level where $P_\mathrm{rege}$ reaches its peak according to \fref{Prege-damping} does not necessarily achieve higher energy efficiency (lower net energy cost) as compared to the regenerative braking scheme. It seems that, although it may regenerate more energy, the inability to vary the damping demands a higher energy movement to fulfil the task.
%\mh{\atFW: Looking at \fref{toy_example}\ref{f:toy-E} isn't the net cost actually \emph{lower} for the fixed damping scheme?? It also does not seem to overshoot (\fref{toy_example}\ref{f:toy-q})??}

Overall, it appears that the proposed hybrid scheme offers the good trade-off between task accuracy and energy efficiency.
\mh{\atFW: I am not asking you to do this right now, but it would be interesting to see how these results vary as you vary the relative importance of accuracy and energy in the cost function used with \ILQR.}

%It can be observed that it costs more energy than for the hybrid mode and recaptures less than the regenerative braking mode, indicating that it achieves better movement performance compared to the regenerative braking mode (Scheme 1) with higher damping range that can be optimally exploited at the cost of decreasing energy efficiency (increasing cost and decreasing regeneration at the same time).
%\mh{\atFW: Last sentence is unclear to me.}

\subsection{Optimal reaching with the MACCEPA-VD}\label{s:maccepavd-model}\noindent
%This section presents the experiments design to address the questions: 
%\begin{enumerate}
%	\item how the cost functions influence the shaping of optimal trajectory?
%	\item how much energy efficiency it can contribute by taking energy regeneration into account?
%\end{enumerate}
%To address the first question, we compare the optimization results of different cost criterion on a \VIA\ actuated robot model --- MACCEPA \cite{VanHam2007} with variable damping (\MACCEPAVD) \cite{Radulescu2012}. The model parameters are identified based on a hardware. Modelling of the energy regeneration and corresponding damping torque calculation are incorporated in this robot model. For the second question, numerical simulations were conducted to compare the trajectories optimized with or without regeneration term $J_\mathrm{rege}$.
%
%To evaluate the effect of regenerative braking on energy optimality, a series of numerical experiments are reported here. The experimental methodology is as follows.
%\mh{\done{Justify your choices of experimental materials and methods, including models used, parameters chosen, controls, etc. How do these choices allow you to show what you set out to show? Describe the aims of each of the experiments.}}
%\subsection{Actuator Model}\noindent
To evaluate the proposed scheme on a more realistic variable impedance actuation system, the MACCEPA-VD mechanism is chosen as an example.%\footnote{Although it has relatively low stiffness range among other alternatives, it has been demonstrated as a good platform and a candidate for many applications, such as walking robots \cite{Enoch2015}. It is also a simple and easy-to-build mechanism.}.
\begin{figure}[t]
	\centering
	\includegraphics[width=0.95\linewidth]{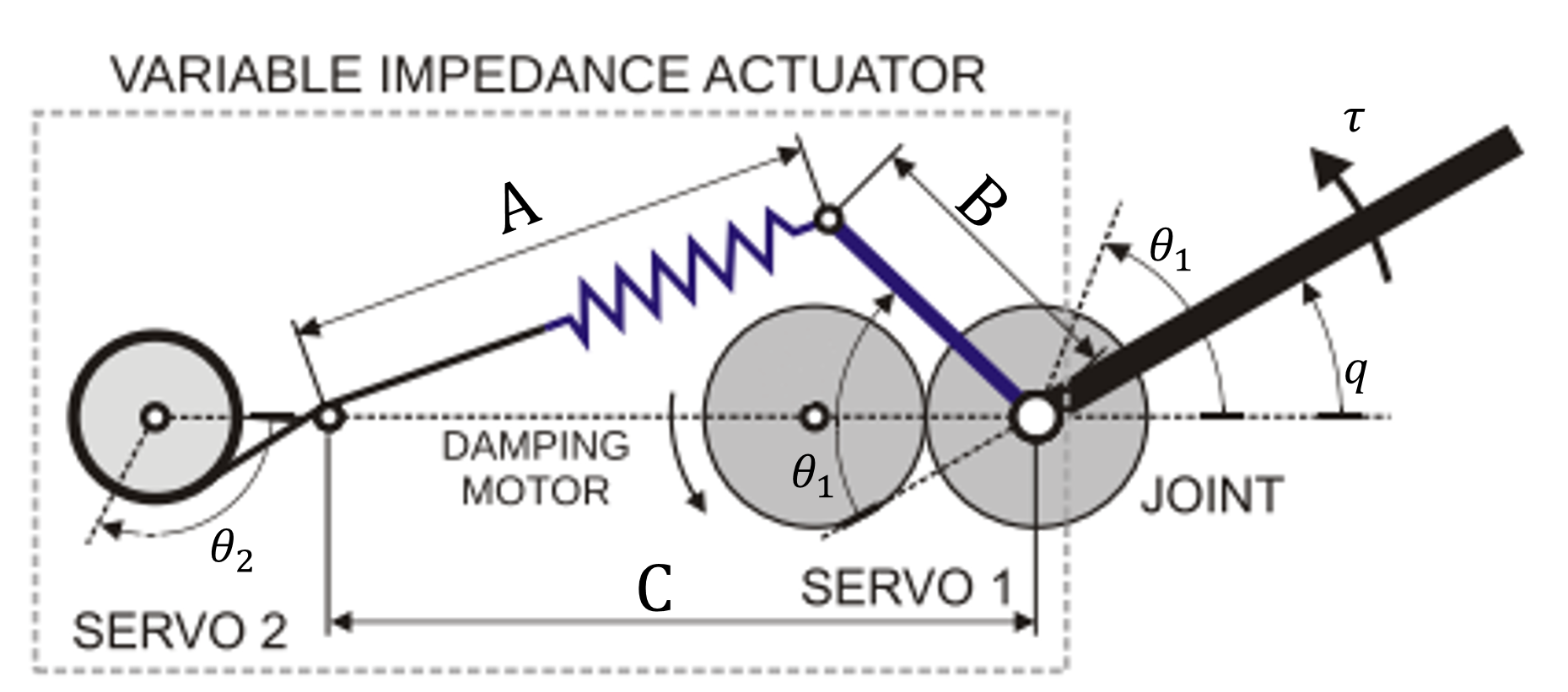}
	\caption{Diagram of \protect\MACCEPA\ \cite{VanHam2007} with variable damping \cite{Radulescu2012}. In the results reported here, $B = 3.6\,\mathrm{cm}, C=13.5 \, \mathrm{cm},r = 1.5 \, \mathrm{cm}$ and the spring has linear spring constant $\kappa=394 \,\mathrm{N/m}$. The link has inertia $m=0.0015 \, \mathrm{kg m^2}$ and friction coefficient $b=0.0023 \, \mathrm{Nms/rad}$.}
	\label{f:maccepavd}
\end{figure}
In the MACCEPA-VD, the equilibrium position and joint stiffness are  controlled by two servomotors separately, while %The spring length during movement depends on the joint state and motor angles. 
the damping coefficient is modulated by changing the duty cycle of the circuit of a DC motor attached rigidly to the joint. The system is illustrated in \fref{maccepavd}.

The forward dynamics for this single joint system can be written as:
\begin{align}
    \ddot{q}    &= (\tau_s - d(u_3) \qdot - b\qdot -\tau_{\mathrm{ext}}) \inertia^{-1} \\\label{e:motor-dynamics1}
	\ddot{\theta}_1 &= \beta^2(u_1 - \theta_1) - 2\beta\dot{\theta}_1    \\\label{e:motor-dynamics2}
	\ddot{\theta}_2 &= \beta^2(u_2 - \theta_2) - 2\beta\dot{\theta}_2 
\end{align}
where $q, \dot{q}, \ddot{q}$ are the joint angle, velocity and acceleration, respectively, $b$ is the viscous friction coefficient for the joint, $\inertia$ is the link inertia, $\tau_s$ is the torque generated by the spring force, and $\tau_{\mathrm{ext}}$ is the joint torque due to external loading (the following reports results for the case of no external loading, \ie $\tau_{\mathrm{ext}}=0$). 
\mh{\done{\atFW: Do you use $\tau_{\mathrm{ext}}$? What is the value?}}
$\theta_1, \theta_2, \dot{\theta}_1, \dot{\theta}_2, \ddot{\theta}_1, \ddot{\theta}_2$ are the motor angles, velocities and accelerations. The motor angles $\theta_1,\theta_2$ are controlled by $u_1 \in [-\pi/3,\pi/3], u_2 \in [0,\pi/3]$ respectively.
%$\tau_{l_1},\tau_{l_2}$ are load torques on the motor shafts. 
The servomotor dynamics \eref{motor-dynamics1}, \eref{motor-dynamics2} are assumed to behave as a critically damped system, with $\beta=30$.
%fitted from data from a hardware implementation of the device (see \sref{experiment}).
%servo 1, servo 2 and damping motor are controlled by control input $\bu=(u_1,u_2,u_3)^\T$. $q$ is the joint angle, $\theta_1$ and $\theta_2$ are used to denote the position of servo 1 and servo 2. The state vector $\bx = (q, \dot{q}, \theta_1, \theta_2, \dot{\theta_1}, \dot{\theta_2})^\T$ in this case has a dimension of 6 (joint dimension $\dimq=1$ and motor dimension $\dimtheta=2$).

The torque $\tau_s$ can be calculated as follows:
%servo 1, servo 2 and damping motor are controlled by control input $\bu=(u_1,u_2,u_3)^\T$. $q$ is the joint angle, $\theta_1$ and $\theta_2$ are used to denote the position of servo 1 and servo 2. The state vector $\bx = (q, \dot{q}, \theta_1, \theta_2, \dot{\theta_1}, \dot{\theta_2})^\T$ in this case has a dimension of 6 (joint dimension $\dimq=1$ and motor dimension $\dimtheta=2$).
\begin{equation}
	\tau_s = \kappa B C \sin{ (\theta_1 - q) } (1+ \frac{r \theta_2 - |C-B|}{A(q,\theta_1)}) 
	%\tau_{l_1} & = \tau_s  \\
	%\tau_{l_2} & = \kappa (r \theta_2 - |C-B| + A(q,\theta_1))
\end{equation}
\mh{\done{\atFW: As far as I can see, $\tau_{l_1}$, $\tau_{l_2}$ are not used or mentioned elsewhere in the paper, so you can omit these to save space.}}
where $A(q,\theta_1) =\sqrt{B^2+C^2-2BC\cos{ (\theta_1 - q) }}$, $B$ and $C$ are the lengths shown in \fref{maccepavd}, $r$ is the radius of the winding drum used to adjust the spring pre-tension, and $\kappa$ is the linear spring constant. \mh{\done{\atFW: Give values for the $B,C,r,\kappa$ used.}}
%\fw{consider move the introduction of MACCEPA model to appendix?}
%\begin{figure}[t]
%	\centering
%	\includegraphics[width=\linewidth]{figures/maccepa_example_new.png}
%	\caption{Test of reaching task on MACCEPA.}
%	\label{f:maccepa_example_}
%\end{figure}
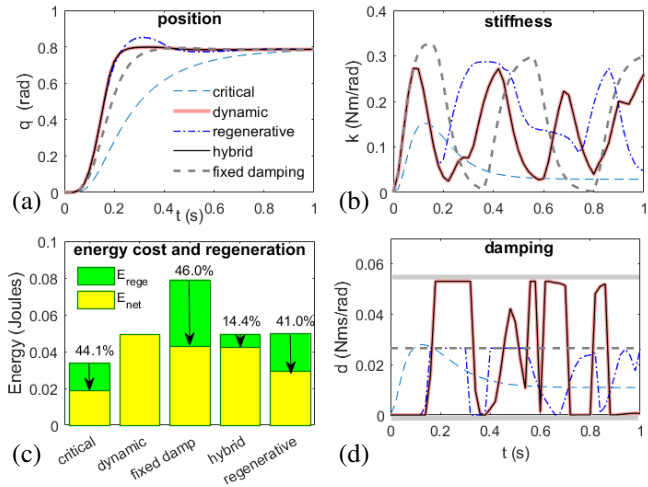
\begin{figure}[t]
	\centering
\begin{overpic}[width=\linewidth]{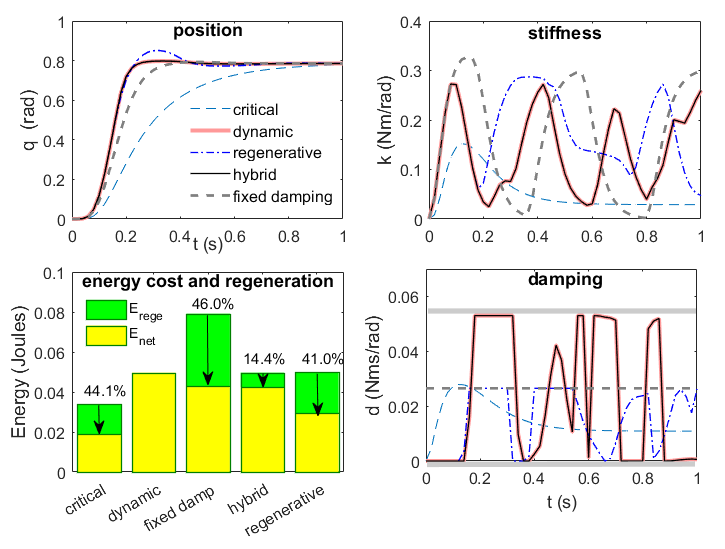}
\put( 2,43){\ref{f:maccepa-q}}
\put(51,43){\ref{f:maccepa-k}}
\put( 2, 5){\ref{f:maccepa-E}}
\put(51, 5){\ref{f:maccepa-d}}
\end{overpic}%    
	\caption{\label{f:maccepa_example} Test of reaching task with the MACCEPA-VD. Shown are optimal
\begin{enumerate*}[label=(\alph*)]
\item\label{f:maccepa-q} joint angular trajectories,
\item\label{f:maccepa-k} stiffness, and
\item\label{f:maccepa-E} total mechanical work and percentage ratio of energy regeneration for different damping schemes, and
\item\label{f:maccepa-d} damping profiles.
\end{enumerate*}
    }
\end{figure}
%\fw{figure needs modification}
\mh{\done{Can we improve the presentation of the energy results in \fref{maccepa_example}, for example, allowing a direct comparison of the energy efficiency?}}
\mh{\atFW: Comparing \fref{maccepa_example} and \fref{toy_example}, it now seems that in the former, the results are more in line with the story you wish to tell  \ie the hybrid scheme achieves lower net energy consumption, without overshoot (unlike \fref{toy_example}, where constant damping is best). Do you have a theory as to why this is? Something to do with the non-linearity of the MACCEPA-VD? You will need to explain this carefully. Looking at \fref{maccepa_example_}, I think you should omit the ``swing back'' example - the results do not look any good (unless I'm missing something, the hybrid approach does no better than dynamic braking here?)}

The damping coefficient $d(u_3)$ depends on control input $u_3$ and is
calculated according to the damping scheme used (\ie \eref{d-scheme-dynamic}, \eref{d-scheme-rege} or \eref{d-scheme-hybrid}). Note also that,
the stiffness of this system depends on the joint and motor positions $q,\theta_1,\theta_2$
\begin{multline}
	k(q,\theta_1,\theta_2) =\kappa BC \cos (\theta_1-q) (1+ \frac{r \theta_2
 - |C-B|}{A}) \\
- \frac{\kappa B^2 C^2 \sin^2 (\theta_1 - q (r \theta_2 - |C-B|))}{A^{\frac{3}{2}}}. \label{e:maccepa-stiffness}
\end{multline}

To evaluate the proposed hybrid damping method, 
%two numerical experiments were designed 
\ILQR\ is used to determine the optimal open-loop control sequence for the task of reaching a target point $q = \pi/4\,\mathrm{rad}$ starting from initial state\footnote{Following \cite{Radulescu2012}, the state vector is defined as $\bx = (q, \dot{q}, \theta_1, \theta_2, \dot{\theta}_1, \dot{\theta}_2)^\T$.} $\bx_0=(0,0,0,0,0,0)^\T$ within finite time $t_f = 2 \, \mathrm{s}$,
%\mh{\atFW: Units for $t_f$.} 
using the proposed hybrid damping scheme (see \sref{hybrid-damping}). The cost function for optimisation takes the same form as \eref{cost-function}, where weighting parameters are $w_1=10^3,w_2=w_3=10^{-4},w_4=10^{-6}$.
\fw{I set the limit of $u_3$ to be 0.8 to make it different from the dynamic.} 
\mh{\atFW: I don't understand this comment.}
\mh{\done{\atFW: What cost function is used?}}

For comparison, the experiment is repeated using dynamic and regenerative braking (Schemes \ref{scheme:dynamic-braking} and \ref{scheme:regenerative-braking}), a fixed damping coefficient of $d=\bar{d}_1$, and a `critically damped' system in which, following \cite{Radulescu2012}, the instantaneous damping ratio is held at $\zeta(t)=1$ by enforcing the stiffness-damping relationship $d(t) = 2\sqrt{k(t)m}$. %as stiffness $k$ varies depending on $q$ according to \eref{maccepa-stiffness}.
\mh{\done{\atFW: How is the 'critically damped' system set up, considering that $k$ depends on $q$? What were the parameters used?}}

The results are shown in \fref{maccepa_example}. There, it can be seen that, the `critically damped' system avoids overshoot, but reaches the target slowly. It also has the lowest energy consumption, in part due to its sluggish response in moving to the target. 
%The fixed damping trajectory in this case has the greatest overshot and oscillation, suggesting that fixing the damping coefficient at $d=\bar{d}_1$ has the worst movement performance. 
The fixed damping trajectory reaches the target faster, but slower than the trajectories using dynamic and hybrid schemes. Although it regenerates the most energy, it also incurs the highest cost in terms of mechanical energy, so overall the net energy cost is higher than Schemes \ref{scheme:regenerative-braking} and \ref{scheme:hybrid-braking} (see \fref{maccepa_example}\ref{f:maccepa-E}). 

The regenerative braking scheme has better energy efficiency as its net energy cost is lower
%greater than the critical damped one, but less than the others.
%\mh{\atFW: How can this be seen?} 
but it suffers greatest overshoot due to its restricted damping range, while the dynamic braking scheme achieves higher accuracy but at the expense of higher net energy cost. The former has higher total mechanical work, but regenerates 41.0\% and results in lower net energy cost. In comparison, hybrid braking achieves almost identical performance in terms of accuracy, but at higher energy efficiency. 
%\mh{\atFW: Did we learn anything new or unexpected from this evaluation? It seems somewhat redundant...}

%However, it should be noticed that, both examples demonstrates the limit of movement performance posed by regenerative braking, but not necessarily suggests that the regenerative braking scheme 1 has overall better energy efficiency. Because the trajectories were optimised with respect to a cost function not emphasising much on decreasing energy cost, so it is natural to suppose that by exploring the weighting parameters of the cost function, it can balance the trade-off between movement performance cost and energy cost according to user needs.

\section{Experiment}\label{s:experiment}\noindent
\begin{figure}[t]
	\centering
	\begin{overpic}[width=\linewidth]{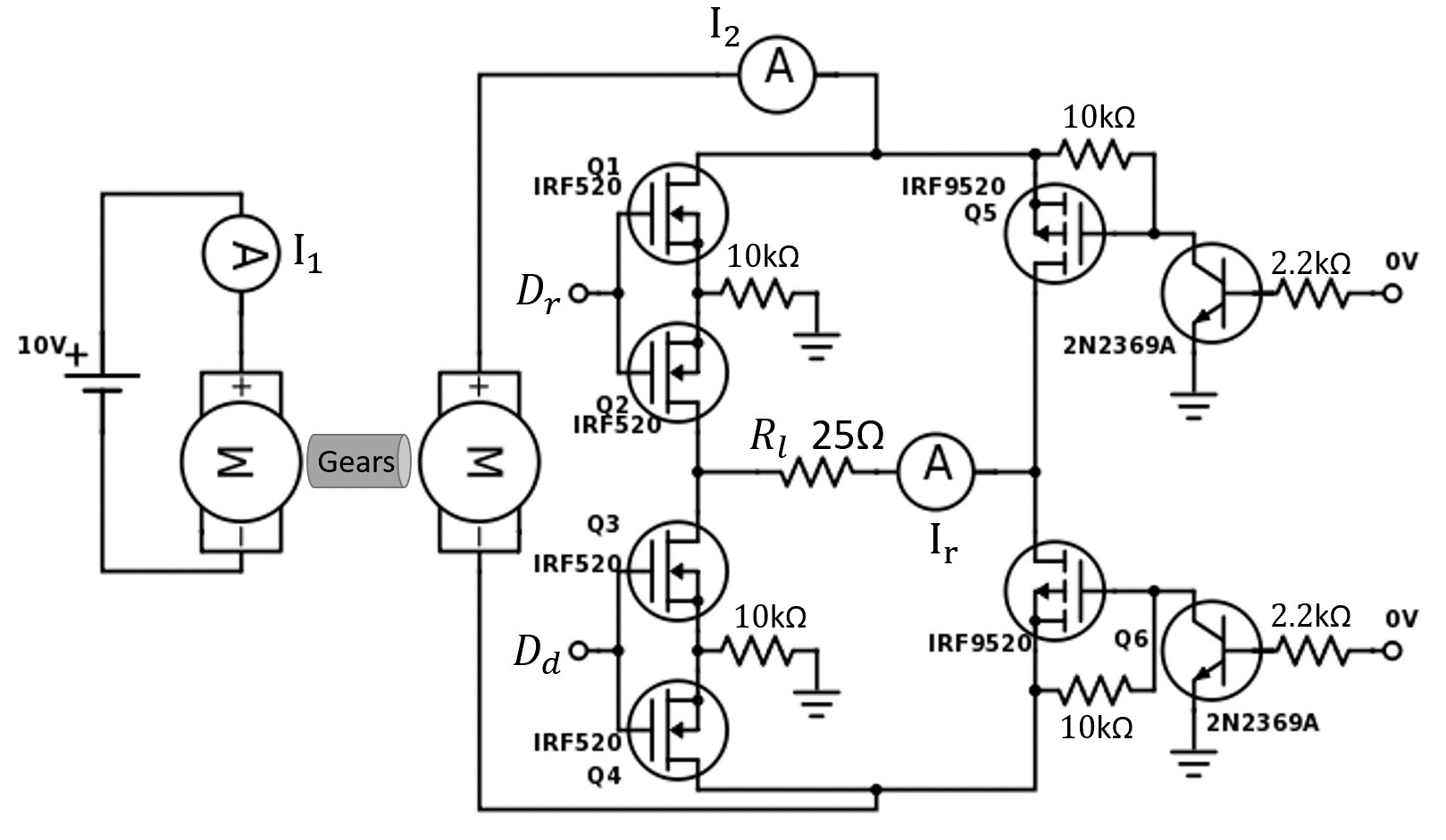}\put(0,5){\ref{f:hw-circuit}}
	\end{overpic}	
\begin{overpic}[width=0.6\linewidth]{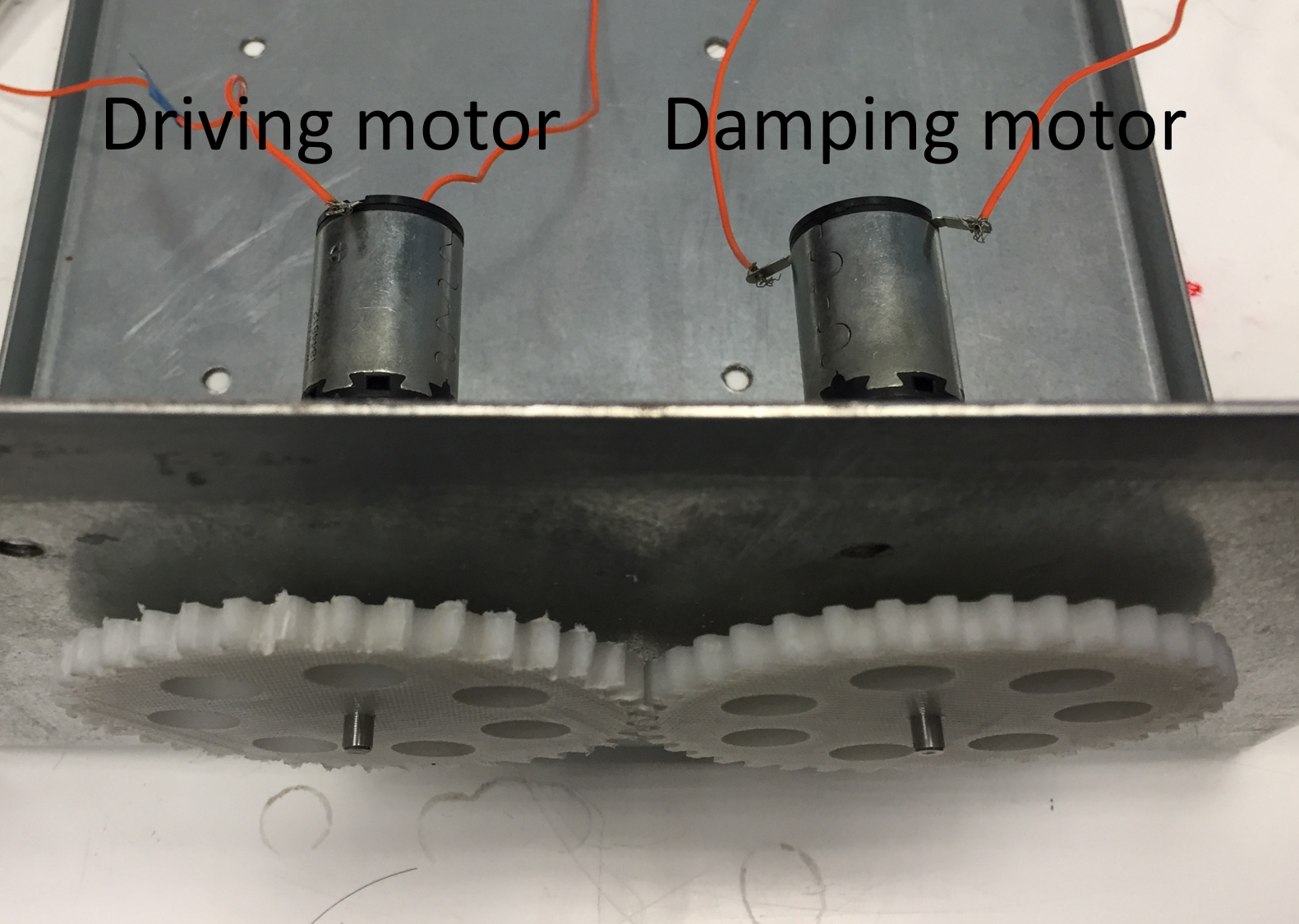}
	\put(0,5){\ref{f:hw-motors}}
\end{overpic}
\begin{overpic}[width=0.38\linewidth]{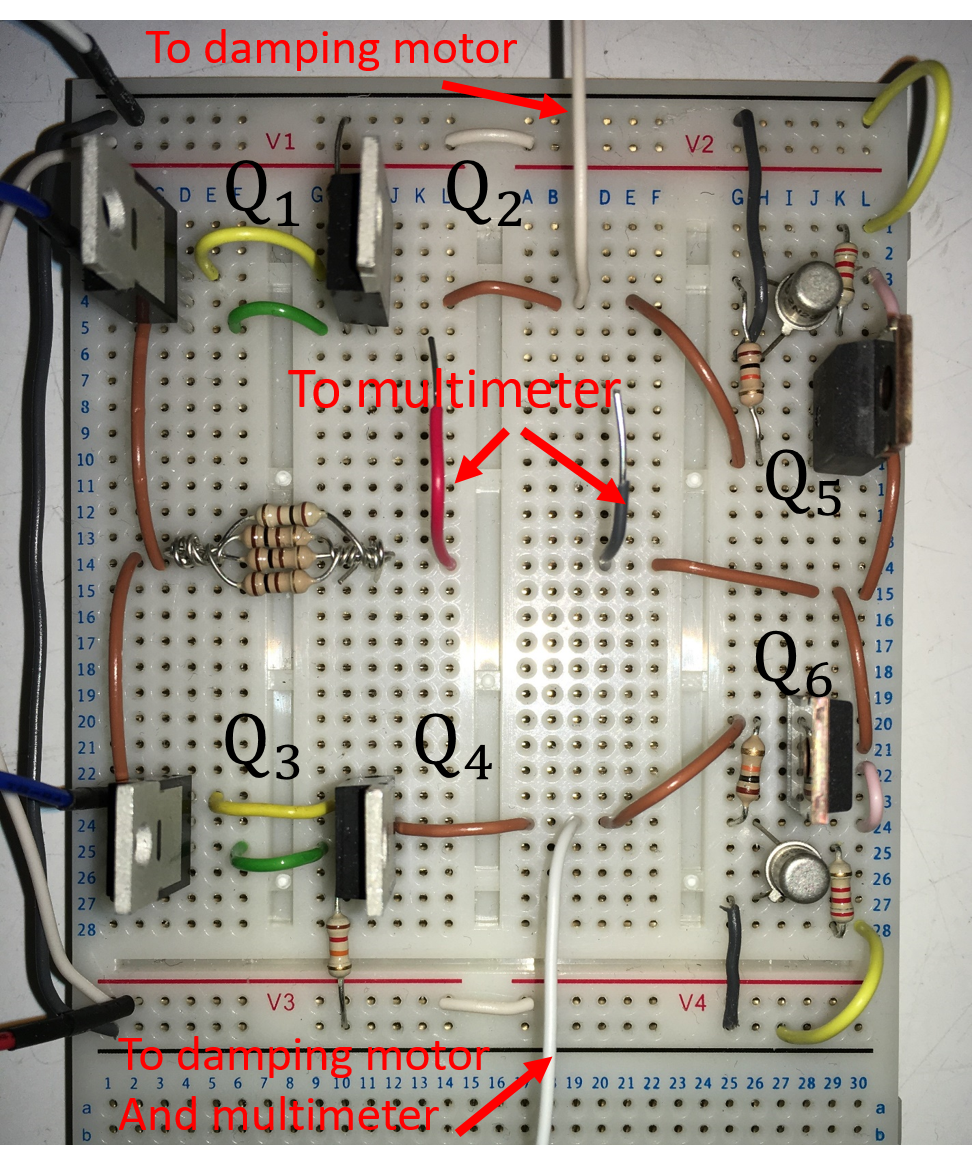}\put(0,5){\ref{f:hw-realcircuit}}
\end{overpic}
\caption{\label{f:hardware-setup}
	Damping and power regeneration measurement experiment setup. Shown are
\begin{enumerate*}[label=(\alph*)]
	\item\label{f:hw-circuit} circuit diagram of the experiment setup,
	\item\label{f:hw-motors} damping test rig, and
	\item\label{f:hw-realcircuit} damping circuit module.
\end{enumerate*}
}
\end{figure}
\mh{\done{\atFW: Some results for \sref{experiment} needed.}}
\mh{\done{\atFW: Please use the \texttt{overpic} package to label the subfigures in \fref{hardware-setup} (see \fref{toy_example} for usage).}}
This section presents an experiment using a physical device to verify the design of the hybrid damping circuit and its control scheme. 

The experimental set up is shown in \fref{hardware-setup}. As a simple test-rig, two identical DC motors
(Maxon A-max 22/110125) are coupled through a pair of spur gears to enable one motor (driver) to drive the other (damper), see \fref{hardware-setup}\ref{f:hw-motors}. The two motors have the same gearhead with $n_d=20$. The torque constant is $k_t = 0.0212 \mathrm{Nm/A}$ and the motor resistance $R_m=21.2\mathrm{\Omega}$.

The damper motor is connected to the circuit depicted in \fref{hardware-setup}\ref{f:hw-circuit}, that is the physical realisation of the conceptual diagram \fref{scheme-bidirection}. In this circuit design, a pair of N-channel MOSFETs (IRF520) is used as one switch to make sure that the switching mechanism works properly for bidirectional current. In \fref{hardware-setup}\ref{f:hw-circuit}, the pair of $Q_1,Q_2$ works as the switch $S_1$, and $Q_3,Q_4$ make the switch $S_2$. Two P-channel MOSFETs (IRF9520) with BJTs (2N2369A) are used as switches $S_3,S_4$. The duty-cycles $D_r,D_d$ are controlled by PWM signals from an Arduino Mega2560 board. By setting $0V$ signals on the control pins for $Q_5,Q_6$, they are open for just one current direction but closed for the other. For the ease of power measurement, a resistor is used to represent the electrical load ($R_l = 25 \mathrm{\Omega}$).

In the experiment, the driving motor is used to drive the system while the damping applied by the second motor is varied, and the resultant motion (motor speeds and energy regeneration) is recorded. Specifically, the driving motor is powered by a 10V DC power supply ($V_{bb} = 10 \mathrm{V}$) while the damping motor control input $u$ is varied from 0 to 1 in increments of 0.1 (with the corresponding duty-cycles $\DCr,\DCd$ computed by \eref{D1D2}). Simultaneously, three multimeters (Rapid DMM 318) are used to measure the currents $I_1,I_2,I_r$ through the driving motor, damping motor and the electrical load $R_l$ respectively. The latter data are used to compute the angular speed of the motors $\omega$ and the damping torque $\tau_d$ according to
\begin{align}
	V_{bb} &= I_1 R_m + n_d k_t \omega \\
	\tau_d &= n_d k_t I_2 = d(u) \omega
\end{align}
The damping coefficient $d(u)$ for a given $u$ is then
\begin{equation}
	d(u) = \frac{n_d^2 k_t^2 I_2}{V_{bb} - I_1 R_m}
\end{equation}
and the regeneration power (normalised by the square of speed for comparison) is estimated as
\begin{equation}
	P_\mathrm{rege} = \frac{n_d^2 k_t^2 I_r^2 R_l}{ (V_{bb} - I_1 R_m)^2 }.
\end{equation}
The results based on the data collected from 10 repetitions of the experiment is plotted in \fref{circuit_exp_result}. 
\begin{figure}
\centering
\includegraphics[width=\linewidth]{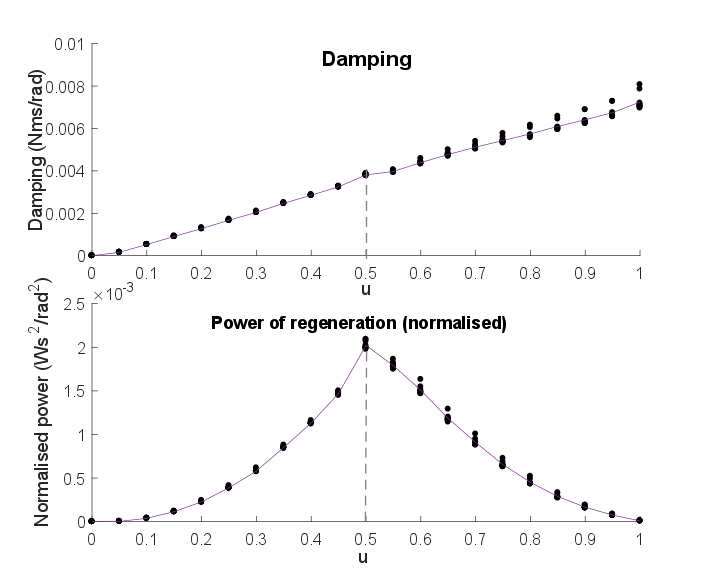}
\caption{Results of the damping test experiment. The damping coefficient (top) and regeneration power (bottom) for each tested control input $u \in [0,1]$ are shown. Black dots represent the data points for 10 repetitions of the experiment and the curves show the mean values.}
\label{f:circuit_exp_result}
\end{figure}

There it can be seen that, the experimental data is in good agreement with the theoretical predictions (see \sref{control-modes}). By increasing $u$ from 0 to 1, the damping coefficient $d$ increases almost proportionally. Furthermore, it is verified that, when fixing the angular speed ($P_\mathrm{rege}$ has been normalised), the relation between $P_\mathrm{rege}$ and $u$ is non-monotonic with a peak found at $u = 0.5$. 

\section{Conclusions}\label{s:conclusion}\noindent
%In this paper we proposed to use energy cost estimated on the motor level as a cost criterion for optimisation. To address the energy efficiency property of VIA, we considered the use of energy regeneration to compensate the energy loss due to damping. It was verified that including power consumption on motor level in cost criterion reduces the energy cost significantly by evaluating the energy-optimal trajectories affected by varying weight parameter $w$ of the energy cost term. Furthermore, it was shown that introduction of energy regeneration on damping device provides possibility to compensate the accuracy-energy trade-off. 
This paper proposes an extension to variable damping module design for VIAs based on the motor braking effect. In contrast to previous, pure dynamic braking designs, the proposed approach provides a solution for realising controllable damping, which enables the VIA to regenerate dissipated energy from bidirectional rotation movement to charge a unidirectional electric storage element. Furthermore, it overcomes the drawback of a reduction in the maximum damping effect found in pure regenerative braking schemes. 

The control input for this damping module simply varies from 0 to 1, representing a proportional percentage of the maximum damping. As the power regeneration has a non-monotonic relation with the control input and damping coefficient (as verified by experiment), the balancing between damping allocation and energy regeneration needs to be treated with care. However, application of the hybrid damping module to VIAs in simulation, shows that it offers more flexibility to balance the trade-off between task performance and energy cost.

In future work, it is intended to investigate the role of variable damping and energy regeneration by considering more use cases and more performance criterion such as reaching time, stability and robustness. Analysis of transmission efficiency of the regenerative damping module will be conducted, by taking account of more factors causing losses during cross-domain energy conversion. Furthermore, it is planned to realise a hardware implementation that integrates the hybrid damping module into a sensorised VIA to measure the combined energy consumption and regeneration during operation. Such information can be used for optimal decision making for task performance and energy cost trade-off. %How the weighting parameter in composite cost function representing the relative importance of energy cost against task goal can be optimally selected according to specific task requirement still remains a question. 
\mh{\done{\atFW: We can revisit the conclusion once the rest of the paper is in shape.}}
\mh{\done{\atFW: Please revise the conclusion according to your latest findings.}}
\fw{\done{One sentence added}}
%\todo[inline]{future works}


%%%%%%%%%%%%%%%%%%%%%%%%%%%%%%%%%%%%%%%%%%%%%%%%%%%%%%%%%%%%%%%%%%%%%%%%%%%%%%%%

%%%%%%%%%%%%%%%%%%%%%%%%%%%%%%%%%%%%%%%%%%%%%%%%%%%%%%%%%%%%%%%%%%%%%%%%%%%%%%%%

%%%%%%%%%%%%%%%%%%%%%%%%%%%%%%%%%%%%%%%%%%%%%%%%%%%%%%%%%%%%%%%%%%%%%%%%%%%%%%%%

%%%%%%%%%%%%%%%%%%%%%%%%%%%%%%%%%%%%%%%%%%%%%%%%%%%%%%%%%%%%%%%%%%%%%%%%%%%%%%%%

\bibliography{IEEEabrv,mybib170301}
%\mh{This is rather a lot of references for a conference paper.}
\bibliographystyle{myIEEEtran}

%\appendix
%\section{Appendix}
%\subsection{Derivation of damping coefficient and regeneration power in switching circuit}
%In a circuit switched on and off by a certain high frequency, the current is typically varying but can be assumed to be constant at steady state and estimated by taking average over one switching period $T$. 

\mh{\atFW: Currently, paper is \pageref{last-page} pages long. Note that there is a charge of $\sim\pounds200$ per page for papers longer than 6 pages!}
\label{last-page}
\end{document}